\documentclass[a4paper,twocolumn,english,superscriptaddress]{revtex4}

\usepackage{amsmath}
\usepackage{graphicx}
\makeatletter

\newcommand{\phFac}[1]{\mathrm{e}^{\mathrm{i} #1}}
\newcommand{\phFacneg}[1]{\mathrm{e}^{-\mathrm{i} #1}}
\newcommand{\planew}[2]{\mathrm{e}^{\mathrm{i} \mathbf{#1} \mathbf{#2}}}
\newcommand{\planewneg}[2]{\mathrm{e}^{-\mathrm{i} \mathbf{#1} \mathbf{#2}}}
\newcommand{\blf}[3]{\mathbf{#1}_{#2 \mathbf{#3}}}
\newcommand{\blfs}[3]{#1_{#2 \mathbf{#3}}}

\newcommand{\ve}[1]{\mathbf{#1}}

\newcommand{\scalProd}[2]{\langle #1 | #2 \rangle}
\newcommand{\bra}[1]{\langle #1 |}
\newcommand{\ket}[1]{|#1 \rangle}

\makeatother

\begin{document}

\title{Efficient construction of maximally localized photonic Wannier 
functions:\\ locality criterion and initial conditions}

\author{Tobias Stollenwerk}

\affiliation{Physikalisches Institut, Universit\"{a}t Bonn, Nussallee 12, 53115 Bonn,
Germany}

\author{Dmitry N. Chigrin}

\affiliation{Theoretical Nano-Photonics Group, Institute of High-Frequency and
Communication Technology, Faculty of Electrical, Information and Media
Engineering, University of Wuppertal, Rainer-Gruenter-Str. 21, D-42119
Wuppertal, Germany}

\author{Johann Kroha}

\affiliation{Physikalisches Institut, Universit\"{a}t Bonn, Nussallee 12, 53115 Bonn,
Germany}

\begin{abstract}
Wannier function expansions are well suited for the description of 
photonic-crystal-based defect structures, but constructing
maximally localized Wannier functions by optimizing the phase degree 
of freedom of the Bloch modes is crucial for the efficiency of the approach. 
We systematically analyze different locality criteria for maximally 
localized Wannier functions in two-dimensional square and triangular 
lattice photonic crystals, employing (local) conjugate-gradient 
as well as (global) genetic-algorithm-based, stochastic methods.
Besides the commonly used second moment (SM) locality measure, we introduce 
a new locality measure, namely the integrated modulus (IM) of the Wannier 
function. We show numerically that, in contrast to the SM criterion, 
the IM criterion leads to an optimization problem with a single extremum, 
thus allowing for fast and efficient construction of maximally localized
Wannier functions using local optimization techniques. 
We also present an analytical formula for the initial choice of Bloch phases,
which under certain conditions represents the global maximum of the IM 
criterion and, thus, further increases the optimization efficiency in 
the general case.
\end{abstract}

% \ocis{000.4430, 230.3120, 230.5298, 250.5300.}

\maketitle

\section{\label{sec:Introduction}Introduction}

Photonic crystals (PhCs) remain to attract a considerable attention
of the scientific community due to their unique properties and potential
technological applications \cite{sak}. To a large extent, PhCs
applications are based on the photonic bandgap effect and involve
sophisticated defect structures, cavities and wave guides in the periodic 
crystal host. 
The full dynamics of the electromagnetic field in such structures 
may be studied in principle rigorously by direct numerical 
solution of Maxwell's equations using the finite difference time domain (FDTD) 
method \cite{Taflove2000}. For the calculation of stationary modes, 
however, the numerical effort may be substantially reduced by using the
Galerkin method \cite{Fletcher1984}, i.e., by expanding  
the electromagnetic field in terms of an appropriate orthonormal set
of basis functions which renders the stationary electromagnetic 
wave equation as a discrete matrix eigenvalue problem. In this case the
proper choice of the set of basis functions is crucial in order to obtain an
accurate description while keeping the dimension of the eigenvalue problem
minimal. 

While for extended wave problems an expansion in terms of 
Bloch functions, the eigenmodes of the unperturbed PhC, is natural
\cite{Chigrin2009,Kremers2009a,Kremers2009}, for the description of defect 
structures the use of Wannier functions as a basis set 
\cite{LEUNG1993,Lidorikis98,Albert2000,Albert2002} is in principle superior, 
because Wannier functions may be constructed as being localized in space 
and are still an exact representation of the point symmetry group of the 
host PhC. Effective solutions developed for
the electronic case, namely, maximally localized generalized Wannier
functions \cite{marzari1,marzari2}, have been applied only recently 
to the electromagnetic case \cite{whittaker,busch2}. Since then,
the theory of photonic Wannier functions has been applied to the analysis
of 2D PhC cavities, waveguides \cite{busch2,Busch2003}, waveguide
crossings \cite{Jiao2005} and PhC heterostructures \cite{Istrate2006}.
The generalization of the approach to the case of 2D slab PhCs and
3D PhCs has been also reported in \cite{McGurn2005,Takeda2006}. 
However, the construction of well-localized Wannier functions 
involves a multidimensional optimization problem \cite{WanProb2}, 
with the arbitrary phase of each Bloch mode (defined below) 
as optimizations parameters. Therefore, the practical importance of this
approach has been fairly limited up to now. In the present work we 
systematically analyze this optimization problem employing local as well 
as global optimization procedures. We propose a novel locality measure 
for the Wannier functions which allows for a highly efficient optimization
of the locality of the Wannier function.

Wannier functions are defined
as the Fourier transform of the Bloch modes 
\begin{equation}
\blfs{\ve{B}}{n}{k}(\ve{r})=\phFac{\phi_{n\ve{k}}}\planew{k}{r}\blfs{\ve{u}}{n}{k}(\ve{r})
\label{eq:bloch_def}
\end{equation}
with respect to the wave vector $\ve{k}$,
\begin{equation}
\blfs{\ve{W}}{n}{R}(\ve{r})=\frac{1}{\sqrt{N}}\sum_{\ve{k}\in\text{BZ}}\planewneg{k}{R}\blfs{\ve{B}}{n}{k}(\ve{r}).\label{eq:wan_def}
\end{equation}
The Bloch mode is a solution of the corresponding wave equation in
a periodic medium, where  $\blfs{\ve{u}}{n}{k}(\ve{r})$ is a lattice-periodic
envelope function, and we have explicitly denoted the arbitrary
phase $\phi_{n\ve{k}}$ of the Bloch function, the Bloch phase. 
$n$ is the band index. 
For simplicity, in the present work we do not consider mixing of different 
bands \cite{Jiao2005,john} in  the construction of Wannier functions from 
Bloch modes.
%We assume
%that the PhC is of finite size with $N$ unit-cells and that cyclic boundary
%conditions are applicable. The number of $\ve{k}$-vectors is therefore
%equal to $N$ and $\ve{k}$-vectors are evenly distributed over the
%first Brillouin zone (BZ). 

From the definitions (\ref{eq:bloch_def}), (\ref{eq:wan_def}) it is 
seen that the Wannier functions would 
be $\delta (\ve{r})$-localized in space only if the 
envelopes $\blfs{\ve{u}}{n}{k}(\ve{r})$ were constant. By contrast, 
for arbitrary Bloch phases $\phi_{n\ve{k}}$ the Wannier functions have  
actually a rather large spatial extension due to the oscillatory character of 
the $\blfs{\ve{u}}{n}{k}(\ve{r})$, a feature which becomes more and more 
pronounced for higher band indices $n$. However, the gauge freedom in the 
Bloch phases may be employed to adjust, for each $\ve{k}$, the value of
$\phi_{n\ve{k}}$ so as to construct maximally localized Wannier 
functions with respect to some locality criterion. This constitutes a 
complex, multidimensional optimization problem.
The choice of the locality measure is not unique. The common choice
used in the literature \cite{busch,whittaker,john} is to minimize the second
moment (SM) of the modulus square of the Wannier function 
\cite{marzari1,marzari2}. Unfortunately, it turned out that the SM, 
as a functional of the set of Bloch phases $\phi_{n\ve{k}}$, has
multiple local minima, so that local optimization procedures, like 
the conjugate gradient method, tend not to find the global minimum 
and the locality of the Wannier function optimized in this way
depends sensitively on the initial set of Bloch phases. 
Global optimization procedures are, on the other hand, slow and
computationally exceedingly expensive.

The purpose of this paper is a systematic analysis of different
locality measures using both local and global optimization methods.
In Section \ref{sec:SM} we construct SM optimized maximally localized
Wannier functions, comparing conjugate gradient (local) and 
genetic-algorithm-based (global) methods. 
This analysis leads us in Section \ref{sec:IM} 
to propose a new locality measure, namely the integrated modulus 
square (IM) measure, resulting
in an optimization problem with a single extremum only. This allows one
to use fast and efficient local optimization techniques to construct
maximally localized Wannier functions. In Section \ref{sec:BC} we
show that, if the Bloch modes conform certain conditions, it is possible
to find an optimal set of Bloch phases with respect to the IM locality
measure analytically. Although the required constrains will not be fulfilled
in the general case, the Wannier functions calculated using such an analytical
set of phases show strong tendency towards localizations and can be
used as an efficient starting point for the numerical optimization. In Section
\ref{sec:Applications} using several examples of PhC defect structures,
the quality of the constructed Wannier functions is demonstrated.
Section \ref{sec:Conclusion} concludes the paper.

\section{Second moment optimization\label{sec:SM}}

\subsection{Definitions}\vspace*{-0.25cm}
In what follows, we limit ourselves to the two-dimensional (2D) case,
which is characterized by the periodic dielectric functions $\epsilon(\ve{r})=\epsilon(\ve{r+R})$,
$\forall\ve{R}\in\mathcal{L}$, with $\ve{r}=(x,y)$ denoting a 2D
vector in the $x$-$y$-plane and $\ve{R}$ being a lattice vector
of some 2D lattice $\mathcal{L}$. In this case, the wave equation
for time harmonic TM (transverse magnetic), $E(\ve{r},t)=\phFacneg{\omega t}E(\ve{r})$,
and TE (transverse electric), $H(\ve{r},t)=\phFacneg{\omega t}H(\ve{r})$,
polarization reads, respectively,
\begin{eqnarray}
\mathcal{L}_{E}E(\ve{r})=-\frac{1}{\epsilon(\ve{r})}\nabla^{2}E(\ve{r}) & = & \frac{\omega^{2}}{\mathrm{c}^{2}}E(\ve{r})\label{eq:waveeqn_TM}\\
\mathcal{L}_{H}H(\ve{r})=-\nabla\frac{1}{\epsilon(\ve{r})}\nabla H(\ve{r}) & = & \frac{\omega^{2}}{\mathrm{c}^{2}}H(\ve{r}).\label{eq:waveeqn_TE}\end{eqnarray}
The wave operators $\mathcal{L}_{E}$ and $\mathcal{L}_{H}$ are hermitian
with respect to the corresponding inner products: \begin{equation}
\langle f|g\rangle_E=\int_{V}d^{2}r{f^{*}({\bf r)\epsilon({\bf r)}}}g({\bf r)}\label{eq:innerProd_TM}\end{equation}
\begin{equation}
\langle f|g\rangle_H=\int_{V}d^{2}rf^{*}(\ve{r})g(\ve{r}). \label{eq:innerProd_TE}\end{equation} 
$V$ is the 2D volume of the crystal. It is important to mention,
that the completeness and orthogonality of the Bloch modes $E_{n\ve{k}}(\ve{r})$
and $H_{n\ve{k}}(\ve{r})$ translates into the completeness and orthogonality
of Wannier functions $\blfs{W}{n}{R}(\ve{r})$ with respect to the
corresponding inner product: \begin{equation}
\scalProd{\blfs{W}{n}{R}}{\blfs{W}{n'}{R'}}_{E/H}=\delta_{nn'}\delta_{\ve{RR'}}.\label{eq:orth_WannierTM}\end{equation}
Moreover, the translation property of the Wannier functions, 
\begin{equation}
\blfs{W}{n}{R}(\ve{r})=\blfs{W}{n}{0}(\ve{r-R})\label{eq:wan_translationBehavior},
\end{equation}
follows from the periodicity of the envelope functions $\blfs{u}{n}{k}(\ve{r})$.

In this work four different two-dimensional photonic crystals are
analyzed for both fundamental polarizations. Square (Sq) and triangular
(Tr) lattices of dielectric rods in air (D) and air rods in dielectric
(A) are considered. In what follows we will refer to these systems
as Sq-D, Tr-D, Sq-A and Tr-A, respectively. The radius of rods and the dielectric constant of
dielectric material and air are chosen to be $r_0/a=0.2$, $\epsilon=12$
and $\epsilon=1$, respectively. We adopt the following definition
of the second moment of the Wannier function $\blfs{W}{n}{R}(\ve{r})$:
\begin{equation}
\mathcal{S}_{n}(\{\phi_{n\ve{k}}\})=\bra{\blfs{W}{n}{R}}(\ve{r}-\ve{r}_{0})^{2}\ket{\blfs{W}{n}{R}}_{E/H},\label{eq:def_SM}\end{equation}
with $\ve{r}_{0}$ being the Wannier center. Two positions of the
Wannier center will be analyzed further, (i) in the center of the scatterer
(``on-site'') and (ii) in the geometrical center between four (three)
scatterers in the case of square (triangular) lattices (``between'').
For example, in a square lattice with lattice constant $a$, the Wannier
center can be set to $\ve{r}_{0}=\ve{R}$ (on-site) or $\ve{r}_{0}=\ve{R+T}$
(between), where $\ve{T}=(0.5a,0.5a)$.

\begin{figure}[tb]
\centering \includegraphics[clip,width=0.95\columnwidth]{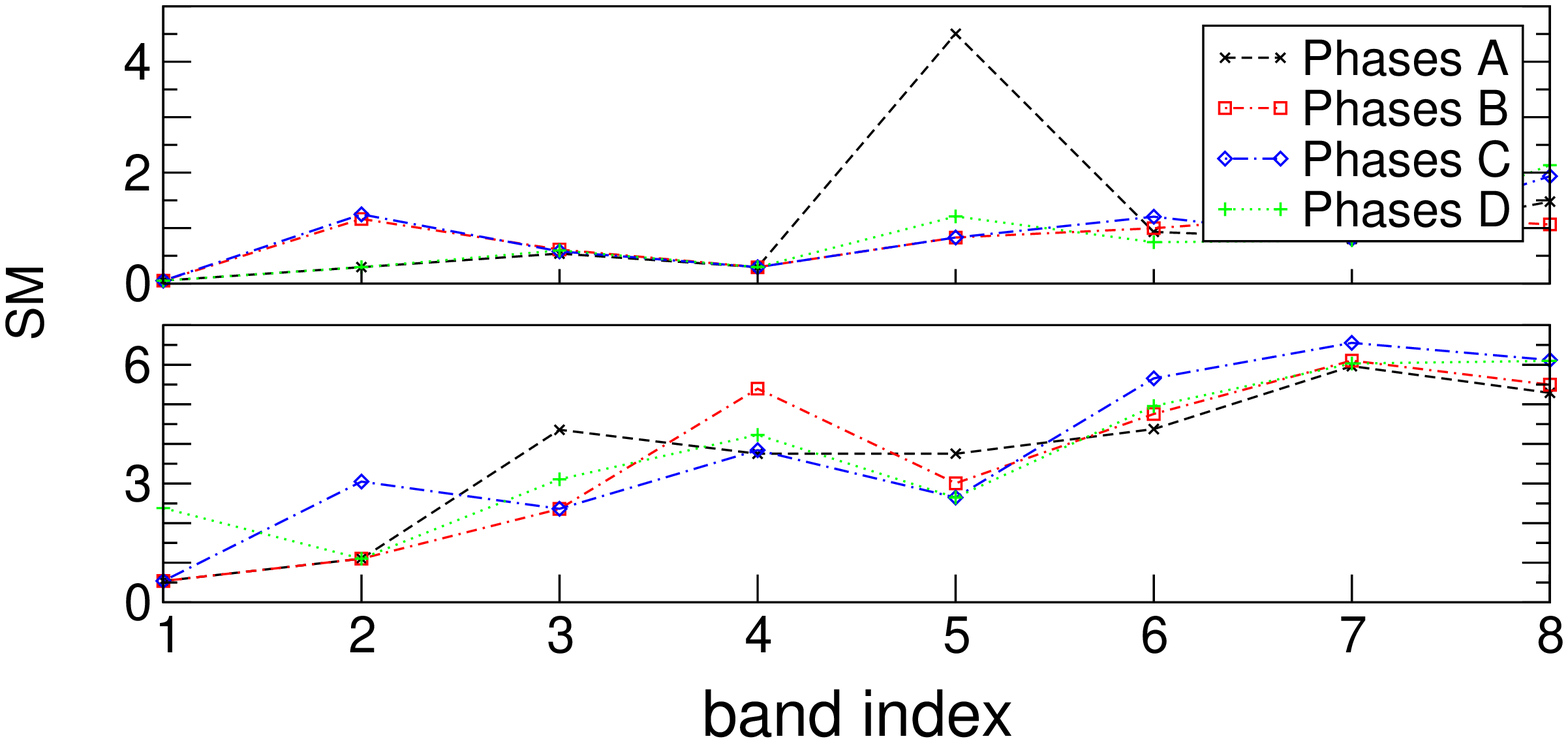} 

\caption[Results of CG-SM optimization]{Second moments $\mathcal{S}_n$
(inverse locality) 
of the Wannier functions in the $n$th band, minimized by using the conjugate
gradient method. Four different, randomly chosen initial sets of Bloch
phases, A, B, C, D, were used for the CG optimization. 
Top: Sq-D crystal, TM polarization. Bottom: Tr-D crystal, TE polarization.}

\label{fig:cg_results_sm} 
\end{figure}

\subsection{Conjugate gradient method}\vspace*{-0.25cm}
The commonly adopted way of finding maximally localized Wannier 
functions is to minimize the corresponding second moment $\mathcal S_n$
with respect to the set of Bloch phases $\phi_{n\ve{k}}$ in the $n$th
band. Note that, by definition of the SM, the regions far away from the
Wannier center (where the Wannier functions usually have an increasingly
complex structure) contribute to the SM with quadratically increasing 
weight. For this optimization we first apply
a standard conjugate gradient (CG) method. This method tends to get
trapped in the local extrema and, hence, requires a careful choice of the initial
set of Bloch phases \cite{john}. In figure~\ref{fig:cg_results_sm}
representative examples of the locality of the SM-optimized Wannier
functions are shown for Sq-D (TM polarization) and Tr-D (TE polarization)
structures. Here and in the following, the Bloch modes have been calculated 
using the plane wave expansion method \cite{mpb-art}. 
The minimized SM (i.e., the inverse locality) of the Wannier functions in the
first eight bands are displayed for four different, random distributions 
of the initial Bloch phases. As expected, the optimal locality obtained using
the CG method depends crucially on the choice of the initial phases. This
is an indication that the SM of the Wannier functions possesses
several local minima.

\begin{figure}[tb]
\centering \includegraphics[clip,width=0.95\columnwidth]{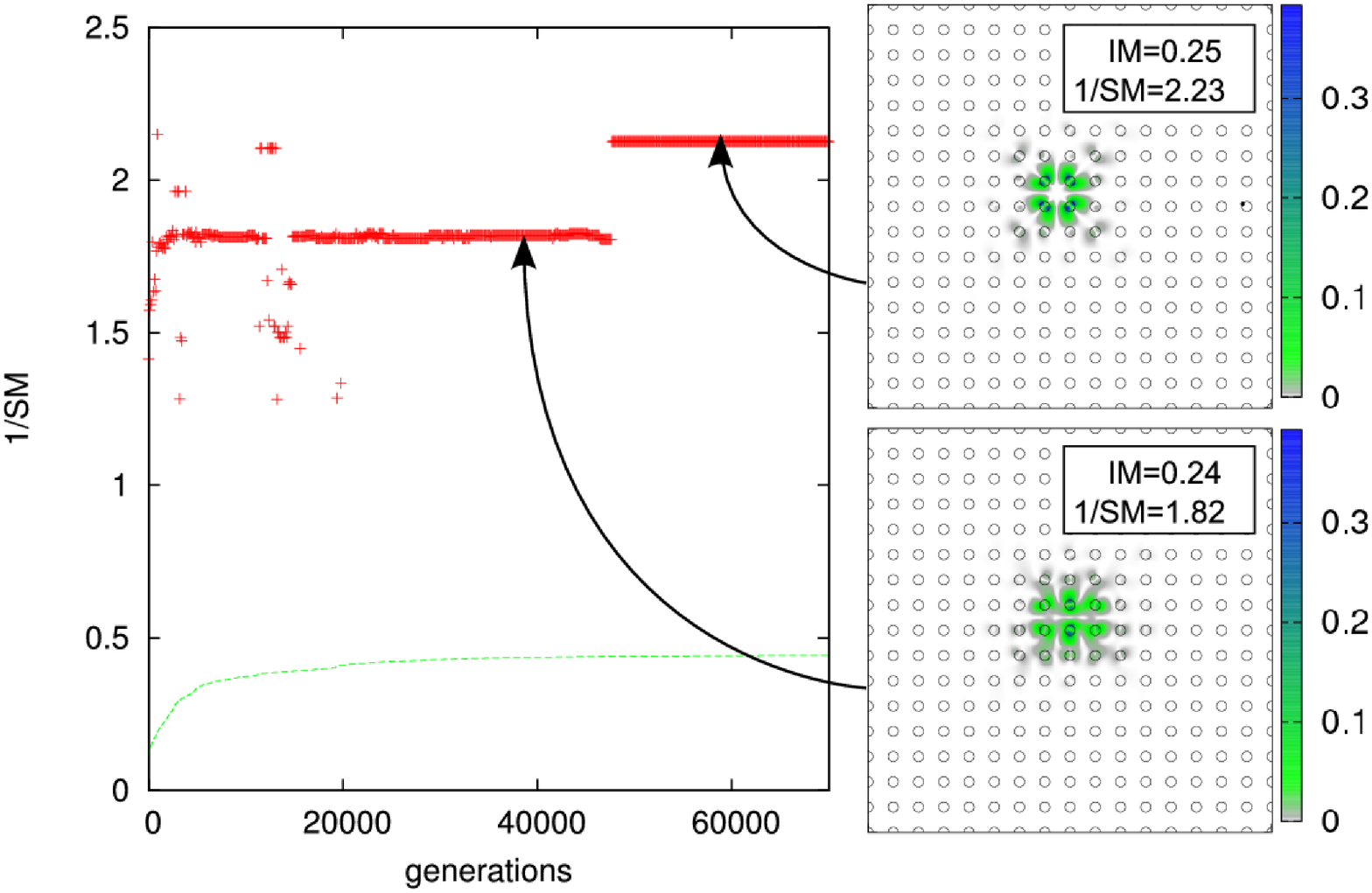}
\\
 \includegraphics[clip,width=0.95\columnwidth]{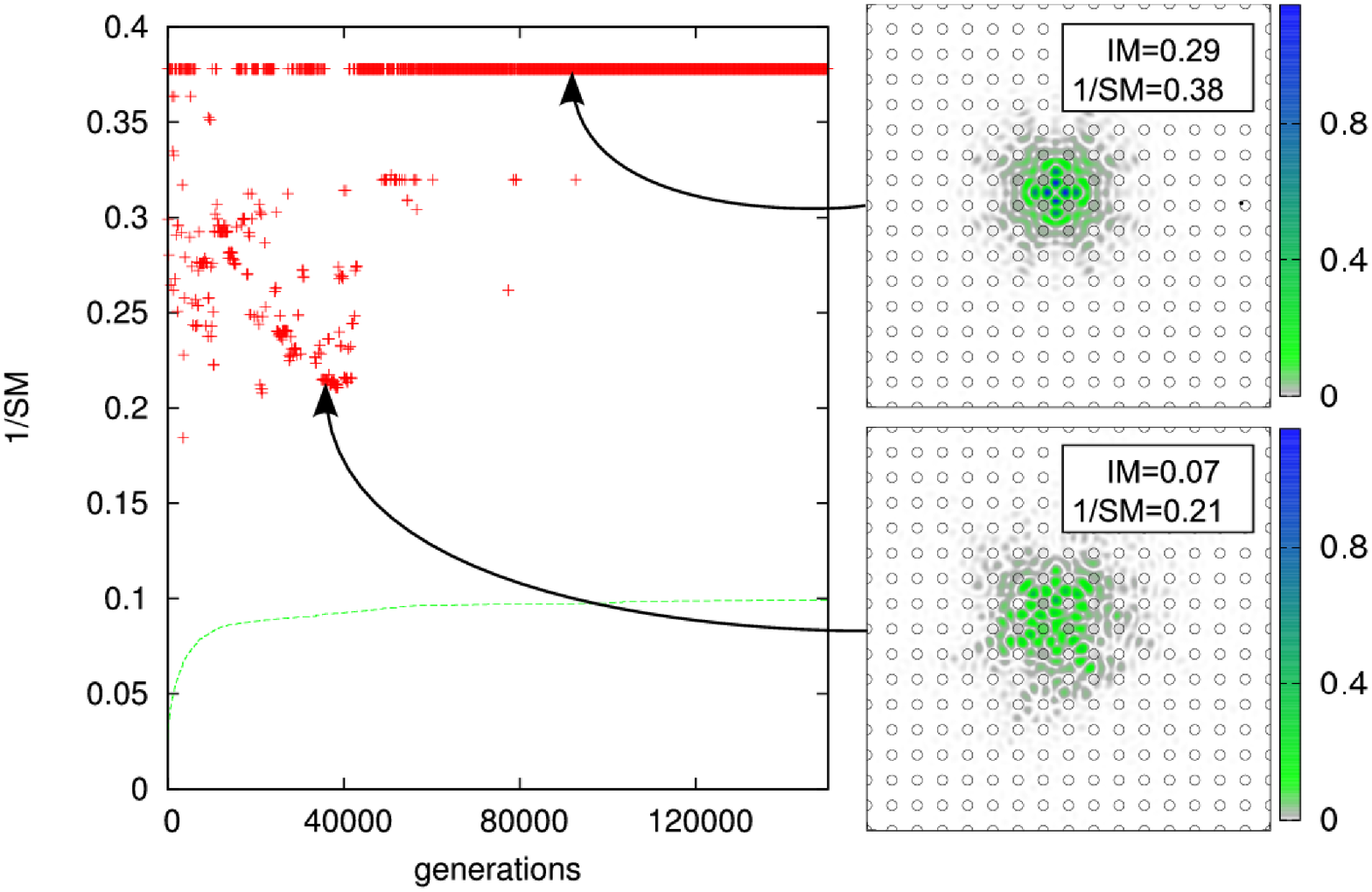} 

\caption[GA evolution for SM]{Locality ($1/\mathcal{S}_{n}$) 
of the SM-optimized Wannier functions
as a function of GA generation for the Sq-D structure, TM polarization,
3rd band (top) and for the Sq-A structure, TE polarization, 5th band (bottom).
The dashed, green line shows the locality of the best-localized Wannier
function in each generation. Every 100 generations, these Wannier functions
served as a starting point for the subsequent CG optimization step
(red crosses). On the right hand side, the modulus square of the
SM-optimized Wannier function is shown for an early (top: 40000th,
bottom: 38300th) and a later (top: 60000th, bottom: 100000th) generation,
respectively.}

\label{fig:ga_sm} 
\end{figure}

\subsection{Genetic algorithm}\vspace*{-0.25cm}
The proper choice of the initial set of Bloch phases is in general
not trivial \cite{john}.
Even if such a choice can be justified, one can never be sure that
the resulting SM-optimized Wannier functions indeed correspond to
the global minimum of the SM. To solve the global optimization problem
and to examine the validity of the SM optimization in more detail, 
we have applied a global,
stochastic-based optimization technique, namely a genetic algorithm
(GA) method \cite{MIT99}. Taking the biologic evolution in nature as a
model, the GA method works with a population of individuals which
pass through a selection procedure and can reproduce themselves. Each
Wannier function represents an individual. The set of Bloch phases,
which determines the Wannier function, is represented as
a large, binary string. The GA method starts with a population
of random Wannier functions and passes them through a selection procedure
where only that one half of the Wannier functions are retained which
are most strongly localized with respect to the given locality criterion
(``most fit individuals). 
In a second step, these survived individuals are allowed to reproduce 
themselves by randomly mixing their strings of phases, thus passing their 
attributes to the offsprings. Together with their off-springs,
the survived individuals, which correspond to better localized Wannier
functions, comprise the new generation. By iterating this procedure over
several thousands of generations the algorithm will converge slowly
but definitely towards the global extremum. Once the GA procedure
has reached the valley of the global extremum, the CG method should 
be applied subsequently to the GA algorithm in order to accelerate the
convergence and improve the accuracy of the solution \cite{MIT99}.

In figure~\ref{fig:ga_sm} the evolution of the GA results is depicted
for two representative systems and polarizations. Every
100 generations a Wannier function with highest locality in the current
population was taken as a starting point for the subsequent CG optimization.
Over the first several thousand generations the locality of the resulting
Wannier functions is varying strongly, indicating hopping of the solution
among different local minima due to the stochastic nature of the algorithm.
At the top panel of figure~\ref{fig:ga_sm} one can observe how the algorithm
is stuck in a local minimum over several thousands of generations, before
it escapes and reaches the global minimum valley at around the 50000th
generation. An improvement of the locality for later generations is
clearly seen from the modulus square of the optimized
Wannier functions (figure~\ref{fig:ga_sm}, right panels). The discontinuous
nature of the GA method ensures with stochastic certainty that the global minimum
of the SM is found, providing the best localization of the Wannier
functions with respect to a given locality measure. At the same time, however,
the numerical load of the GA method exceeds the one of the CG method
by far, making it inappropriate for routine application for an efficient
construction of maximally localized Wannier functions.

\section{Integrated modulus optimization\label{sec:IM}}

\begin{figure}[tb]
\centering \includegraphics[clip,width=0.95\columnwidth]{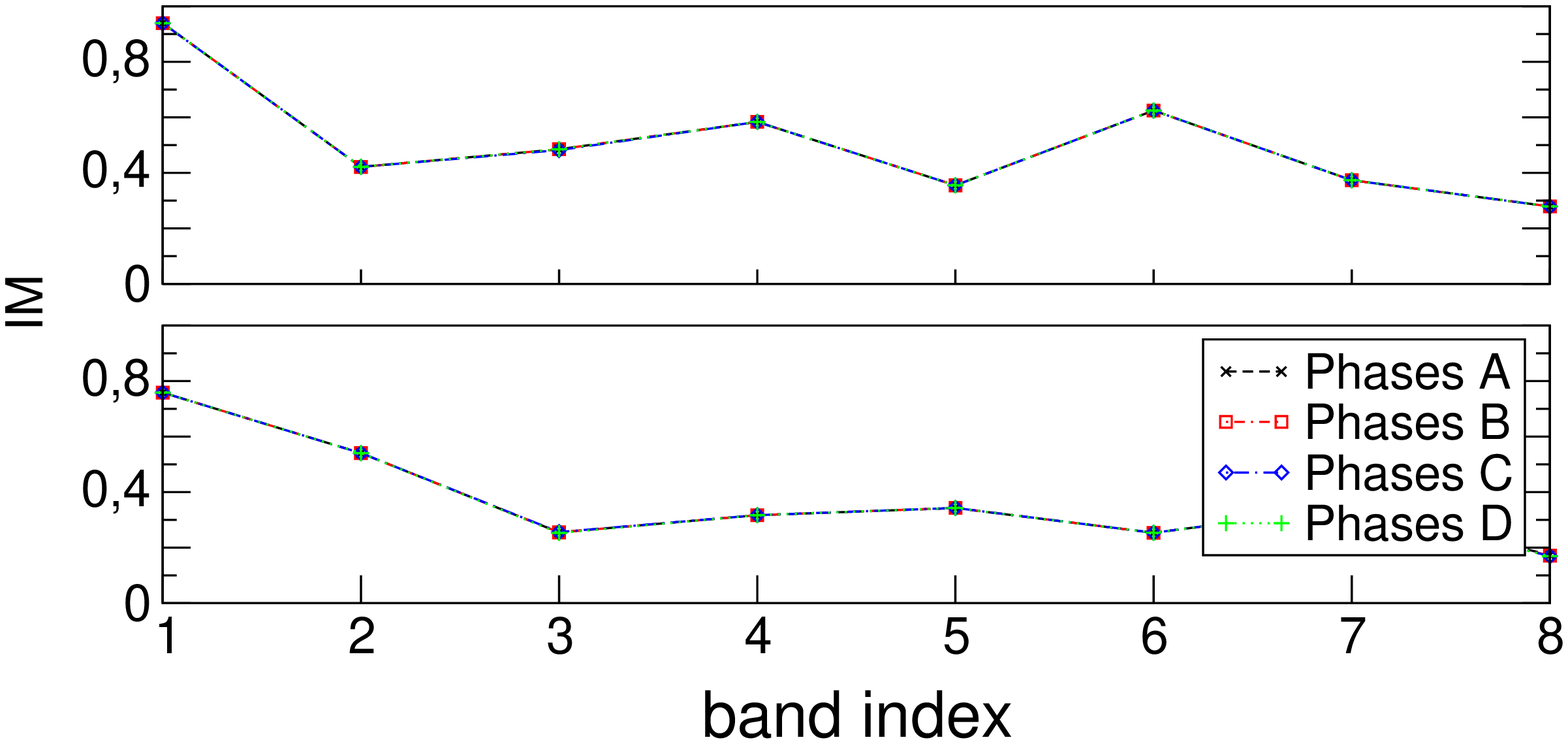} 

\caption[Results of CG-IM optimization]{Locality ($\mathcal{I}_{n}$) 
of the Wannier functions optimized
with respect to the IM locality measure using the CG method. Four
different randomly chosen initial sets of Bloch phases were used for
the CG optimization. Top: Sq-D crystal, TM polarization. Bottom: Tr-D
crystal, TE polarization.}

\label{fig:cg_results_im} 
\end{figure}

The complicated structure of the Wannier functions at large distances,
which is expressed by several local minima of their SMs
and the associated difficulties in the construction
of maximally localized Wannier functions, motivates the search for a
simpler criterion for the locality of Wannier functions. Here, we
introduce a new criterion, the integrated modulus square (IM), defined
as 
\begin{equation}
\mathcal{I}_{n}(\{\phi_{n\ve{k}}\})=\int_{\text{UC}}d^{2}r\blf{W}{n}{R}^{*}(\ve{r})X(\ve{r})\blf{W}{n}{R}(\ve{r}),\label{eq:def_IM}\end{equation}
where the integration region is the first unit cell (UC) around the Wannier
center (UC). We choose the function $X(\ve{r})$ in such a way, that
the IM is equal to unity for a Wannier function which is completely
confined within such a unit cell. i.e., 
\begin{equation}
X(\ve{r})=\left\{ \begin{array}{ll}
\epsilon(\ve{r}) & \quad\mbox{for TM}\\
1 & \quad\mbox{for TE}\end{array}\right. \ . 
\label{eq:IMnormalization}
\end{equation}
A well localized Wannier function corresponds to a large IM, and one
needs to maximize the IM in order to obtain the maximally localized
Wannier functions. The IM is a very sharp criterion, since it does not
depend on the structure of the Wannier functions outside of the integration
region at all.

We examined the IM locality measure as a localization criterion for the same four different physical
systems and both polarizations as it was done in the SM case. To that end we construct
maximally localized Wannier functions using the same four different,
randomly chosen initial sets of Bloch phases as in Section \ref{sec:SM}. Representative examples of
CG optimization are shown in figure~\ref{fig:cg_results_im}. In
contrast to the SM case, the localities of the resulting Wannier functions
coincide for all four sets of phases. This is the case for all considered
systems and polarizations, strongly indicating that the IM locality
measure does not possess any local extrema.

\begin{figure}[tb]
\centering \includegraphics[clip,width=0.95\columnwidth]{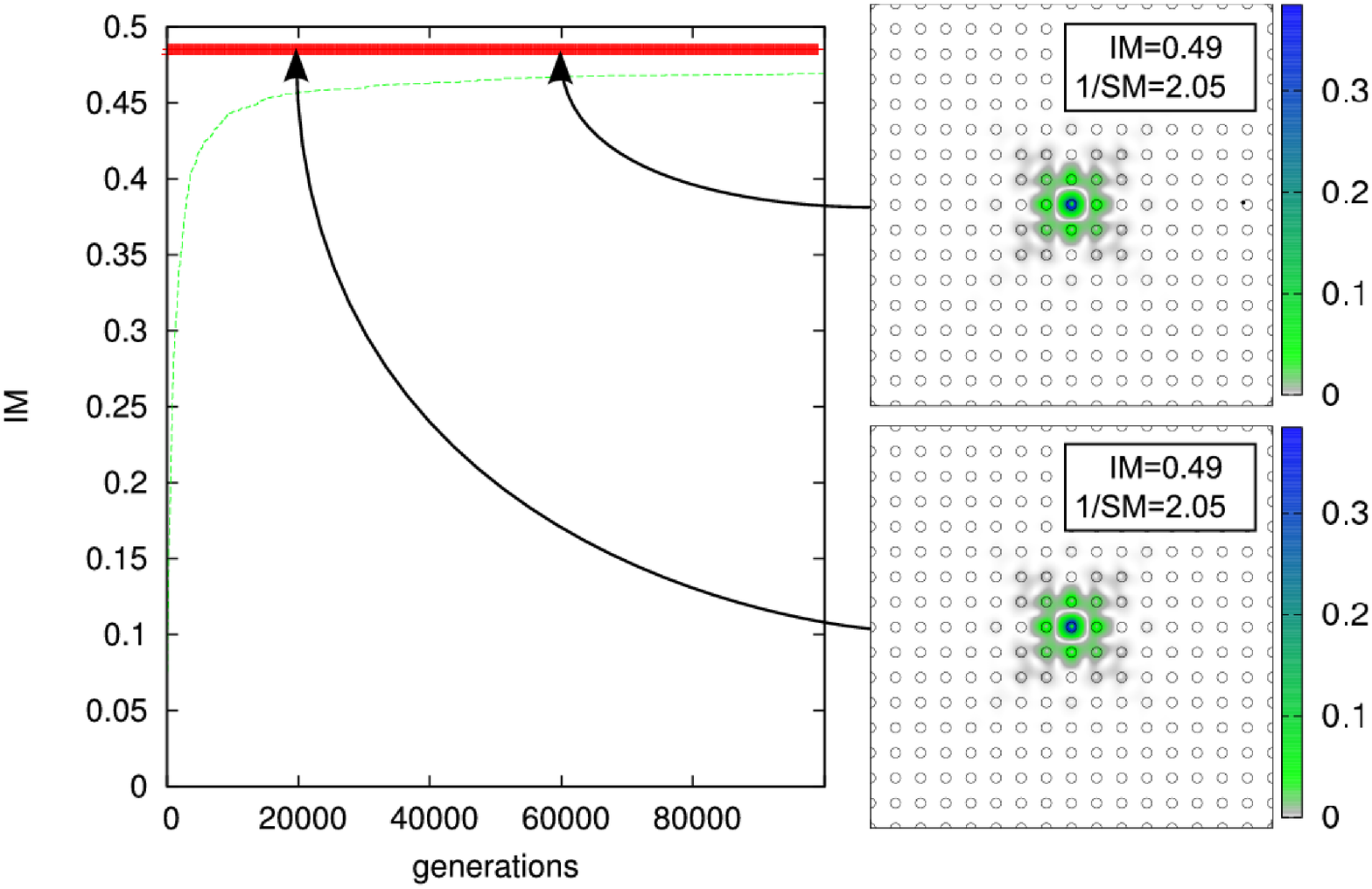}
\\
 \includegraphics[clip,width=0.95\columnwidth]{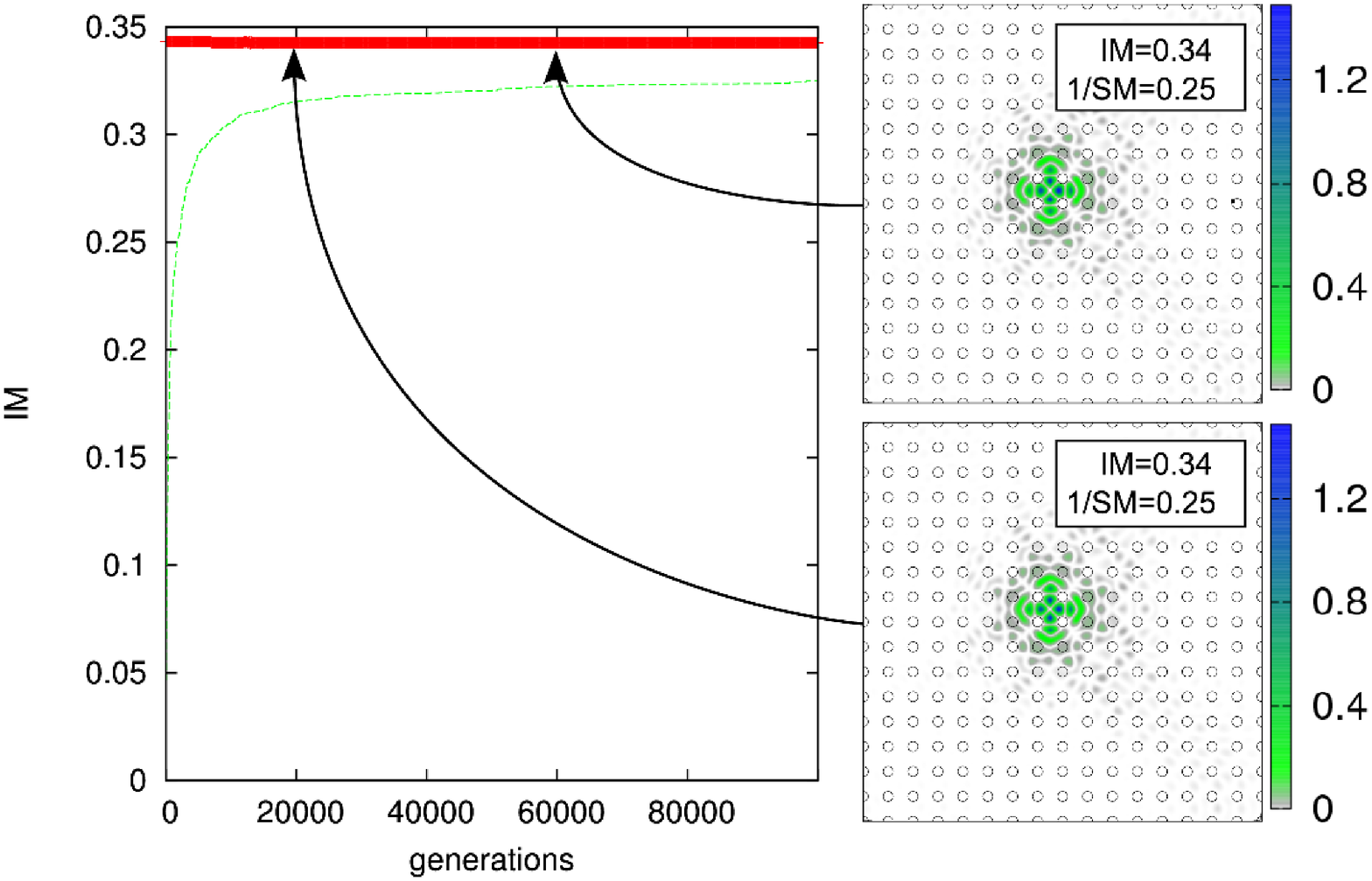} 

\caption[GA evolution for IM]{Locality ($\mathcal{I}_{n}$) of the 
IM-optimized Wannier functions
as a function of GA generation for the Sq-D structure, TM polarization,
3rd band (top) and for the Sq-A structure, TE polarization, 5th band (bottom).
The dashed green line shows the locality of the best-localized Wannier
function in each generation. Every 100 generations these Wannier functions
served as a starting point for the subsequent CG optimization step
(red crosses). On the right-hand side, the modulus square of the
SM-optimized Wannier functions is shown for an early (20000th) and
a later (60000) generation, respectively.} 
\label{fig:ga_im} 
\end{figure}

\begin{figure}[tb]
\centering \includegraphics[width=0.9\columnwidth]{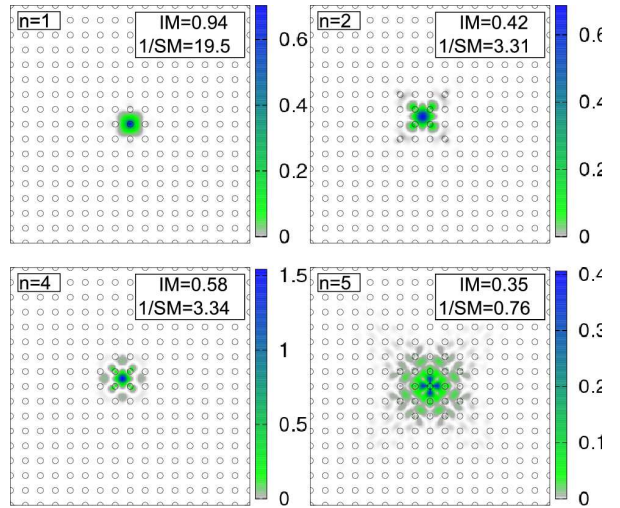} 

\caption[MLWF - Dielectric rods in air, TM]
{Modulus square of the maximally localized Wannier functions (with
respect to the IM) for the Sq-D structure (TM polarization). The Wannier
center was chosen as 'on-site' for the 1st and 5th band and as 'between'
for the 2nd and 4th band.}

\label{fig:MLWF_dielRods_tm} 
\end{figure}

\begin{figure}[tb]
\centering \includegraphics[width=0.95\columnwidth]{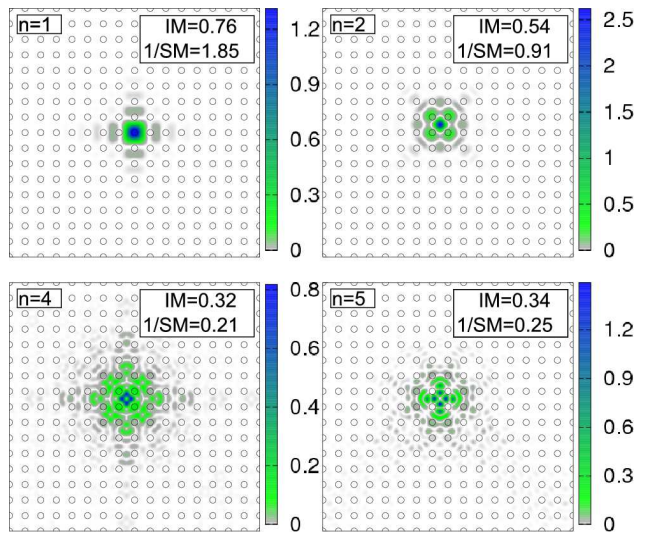} 

\caption[MLWF - Air pores in a dielectric - TE]{ Modulus square of the maximally localized Wannier functions (with
respect to the IM) for Sq-A structure (TE polarization). The Wannier
center was chosen as 'on-site' for the 1st band and as 'between' for
the 2nd, 4th and 5th band. }

\label{fig:MLWF_airRods_te} 
\end{figure}

To support this hypothesis, we applied the GA method to solve the optimization
problem. For all considered structures no significant variation of
the locality has been observed for different GA generations. Representative
examples are shown in figure~\ref{fig:ga_im}. One can clearly see,
that over many thousands of generations the locality of the Wannier
functions optimized with respect to IM criterion stays constant. This
proves numerically that the considered optimization problem possesses a
single extremum, making the procedure independent on the choice of
the initial set of Bloch phases. As a consequence, the use of the
IM as a locality criterion together with the CG as an optimization
method represents a fast as well as reliable method for the construction
of maximally localized Wannier functions. Figures~\ref{fig:MLWF_dielRods_tm}
and \ref{fig:MLWF_airRods_te} show the maximally localized Wannier functions
for the first several bands of the Sq-D (TM polarization) and Sq-A
(TE polarization) structures, respectively. In both cases the IM criterion
was used along with the CG method. The optimized Wannier functions demonstrate
good locality which degrades slowly with increasing band indices,
since the envelope functions become more and more oscillatory, reflecting
the not-plane-wave like nature of Bloch modes. It is worth noting, that
the Wannier functions optimized with respect to the IM and SM criteria
are in general not equal, even if the global minimum of the SM or the
global maximum of the IM has been reached. For example for the system
of dielectric rods in air (TM polarization) the SM- and IM-optimized
Wannier functions coincide for the 1st and the 2nd band (not shown), but not
for the third band (top right Wannier function in
figures~\ref{fig:ga_sm} and \ref{fig:ga_im}).

Sometimes, it is beneficial to use the time-reversal symmetry of wave
operators $\mathcal{L}_{E}$ and $\mathcal{L}_{H}$ to construct real
Wannier functions. This is possible, since the envelope functions of the
Bloch modes transform as 
\begin{equation}
\blfs{u}{n}{-k}(\ve{r})=\blfs{u}{n}{k}^{*}(\ve{r})
\end{equation}
under time reversal (inversion of the reciprocal space), and the Bloch 
phases can be constrained by the condition
\begin{equation}
\phi_{n\ve{k}}=-\phi_{n\ve{-k}}\ .\label{eq:phaseInversion}
\end{equation}
By restricting the
parameter space of the optimization problem in such a way, we found 
that both SM and IM optimization criteria possess multiple extrema, making the
use of local optimization method not efficient without special choice
of the initial Bloch phases. However, using real-valued Wannier functions
is of advantage for the considerations of the next section.

\section{Choice of initial conditions\label{sec:BC}}

Here we propose an analytical expression for a generic set of Bloch phases 
to be used as a starting point for the optimization 
procedure. It is based on the following theorem.
 
{\it Theorem:}
Suppose that 
(i) the Wannier functions are real-valued, i.e., 
$\phi_{n\ve{k}}=-\phi_{n\ve{-k}}$, and
the Bloch functions $\blfs{\ve{B}}{n}{k}(\ve{r})=\phFac{\phi_{n\ve{k}}}
\blfs{\ve{\tilde B}}{n}{k}(\ve{r})$ 
conform to the following conditions:
$\text{Re}(\blfs{\ve{B}}{n}{k}(\ve{r}))$ has 
the same sign (i) for all points $\ve{r}$ in the unit cell and 
(ii) for all wave vectors $\ve{k}$ in the first Brillouin zone
(for each component of the vector Bloch function).
Then maximizing the IM of the Wannier functions,
\begin{equation}
\mathcal{I}^W_{n}(\{\phi_{n\ve{k}}\})=\int_{\text{UC}}d^{2}r
\left|\blf{\ve{W}}{n}{R}(\ve{r})\right|^2\end{equation}
with respect to the Bloch phases $\phi_{n\ve{k}}$ is equivalent to 
maximizing the IM of the real part of the Bloch functions in the first 
unit cell around the Wannier center,
\begin{equation}
\mathcal{I}^B_n(\{\phi_{n\ve{k}}\})=\int_{\text{UC}}d^{2}r\left(\text{Re}(\phFac{\phi_{\ve{k}}}\ve{\tilde{B}}_{n\mathbf{k}})\right)^{2}
\label{eq:blochCrit}\end{equation}
for each wave vector in the first Brillouin zone separately.
This theorem holds in any spatial dimension, even though in this paper 
we consider two-dimensional systems only. 
Note that in the above expressions we have set the weight factor 
$X(\ve{r})$ appearing in Eq.~(\ref{eq:def_IM}) equal to unity for 
simplicity.

The proof is straightforward. 
Due to the translation property of the Wannier function,
\eqref{eq:wan_translationBehavior},
it is sufficient to prove the equivalence for $\ve{R}=\ve{0}$ only.
In this case 
\begin{equation}
\blfs{\ve{W}}{n}{0}(\ve{r})=\frac{1}{\sqrt{N}}\sum_{\ve{k}\in\text{BZ}}
\blfs{\ve{B}}{n}{k}.
\end{equation}
\begin{figure}[tb]
\centering \includegraphics[clip,width=0.95\columnwidth]{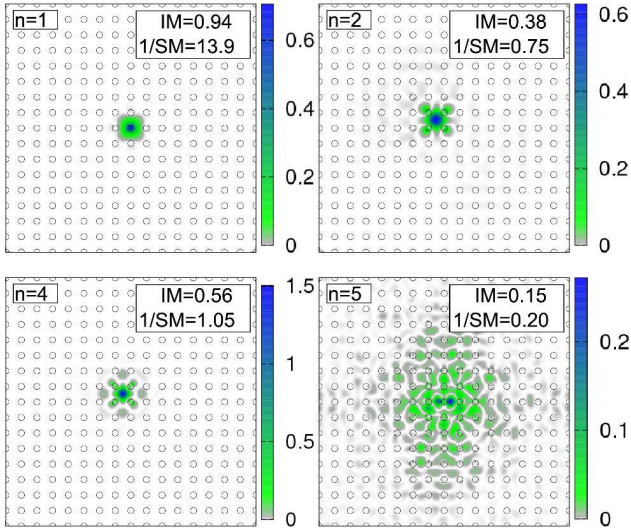} 

\caption[MLWF]{Modulus square of Bloch-criterion (Eq.~\ref{eq:blochCrit}) optimized Wannier 
functions for a square lattice of dielectric rods in air (TM polarization). For
n=1,2,4 the Bloch criterion is equivalent to the IM-criterion for
real Wannier functions. For n=3,5,6 the Wannier functions show at
least a tendency to localize around the Wannier center.}
\label{fig:blochOptWF} 
\end{figure}
It is easy to recognize that the equality
\[
\int_{\text{UC}}d^{2}r\left|\text{Re}(\blfs{\ve{W}}{n}{0}(\ve{r}))\right|=\frac{1}{\sqrt{N}}\sum_{\ve{k}\in\text{BZ}}\int_{\text{UC}}d^{2}r\left|\text{Re}(\blfs{\ve{B}}{n}{k})\right|\]
holds, if the conditions (ii) and (iii) are fulfilled.
Since the Wannier functions are chosen to be real-valued (i), the 
functional on the left-hand side is maximized by the same set of 
$\phi_{n\ve{k}}$ as $\mathcal{I}^W_{n}(\{\phi_{n\ve{k}}\})$, which proves 
the theorem. 

The maximization problem \eqref{eq:blochCrit} can be solved analytically
leading to the following set of Bloch phases 
\begin{equation}
\tan(2\phi_{\ve{k}})=\frac{-\int_{\text{UC}}d^{2}r2\text{Re}(\tilde{B}_{n\mathbf{k}})\text{Im}(\tilde{B}_{n\mathbf{k}})}{\int_{\text{UC}}d^{2}r\{\text{Re}(\tilde{B}_{n\mathbf{k}})^{2}-\text{Im}(\tilde{B}_{n\mathbf{k}})^{2}\}},
\label{eq:blochCrit_solution}\end{equation}
where $\tilde B_{n\ve{k}} = 
\left[\ve{\tilde B}_{n\ve{k}}(\ve{r})\cdot\ve{\tilde B}_{n\ve{k}}(\ve{r})\right]^{1/2}$
is the (complex) amplitude of the vector Bloch function. Relation \eqref{eq:blochCrit_solution}
defines the Bloch optimization criterion. Note that the phases which are transformed 
by $\phi_{\ve{k}}\rightarrow\phi_{\ve{k}}+\pi$,
or equivalently $\phFac{\phi_{\ve{k}}}\rightarrow\phFac{(\phi_{\ve{k}}+\pi)}=-\phFac{\phi_{\ve{k}}}$,
also fulfill \eqref{eq:blochCrit_solution}. Furthermore, the phases
which fulfill \eqref{eq:blochCrit_solution} also obey \eqref{eq:phaseInversion}
and, therefore, the Wannier functions optimized with respect to the
Bloch criterion \eqref{eq:blochCrit} are real-valued, as initially assumed.

Due to the strongly oscillatory nature of Bloch functions in coordinate
space (especially for the higher order bands), condition (ii) will
hardly be fulfilled exactly in a realistic system.
But the conditions (i) and (iii) can always 
be fulfilled simply by correct choice of the phase factor sign. 
In figure \ref{fig:blochOptWF} an example
of the Bloch criterion optimization is shown for the Sq-D structure (TM
polarization). For the 1st, 2nd and 4th bands Bloch functions fulfill
Bloch criterion conditions at least approximately, and an analytical
set of Bloch phases leads to well localized Wannier functions. Even
in the case when the Bloch functions do not maximize the Bloch criterion
Eq.~(\ref{eq:blochCrit})
exactly, the Wannier functions optimized with respect to \eqref{eq:blochCrit}
exhibit a tendency to localize around its Wannier center. The same
tendency has been obtained for all four test structures in both fundamental
polarizations. This suggests to use the analytical set of Bloch phases
\eqref{eq:blochCrit_solution} as an initial set of phases for the numerical
optimization, using either second moment or integrated modulus methods.
This should help avoiding the local minima trapping problem and can
reduce the computation time considerably.

\section{Wannier function quality\label{sec:Applications}}

\begin{figure}[t]
\centering \includegraphics[angle=-90, width=0.95\columnwidth]{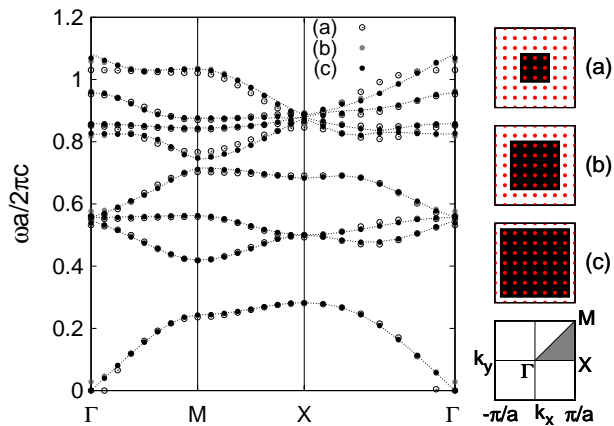} 

\caption[Tight-binding reproduced band structure]{Band structure of a square lattice of dielectric rods in air (TM
polarization). The dashed curve is the original band structure, the
solid dots are the reproduced band structure obtained by within Wannier
function formalism. On the right side the set of next neighboring
sites surrounding Wannier center $\ve{R}=\ve{0}$ is shown for different
nearest neighbor approximations.}

\label{fig:reprBS} 
\end{figure}

\begin{figure}[b]
\centering \includegraphics[width=0.95\columnwidth]{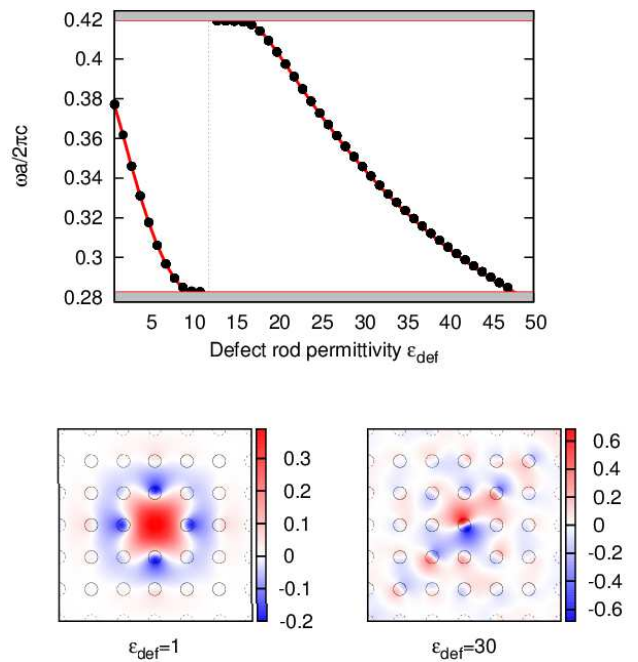} 

\caption[Point defect frequencies and modes]{Frequencies of the modes in a point defect consisting of a single
rod with differing permittivity $\epsilon_{def}$ in a a square lattice
of dielectric rods in air (TM polarization). The dots indicate the
results of the Wannier function approach by taking the first eight
bands into account. They are in complete agreement with plane wave
calculations (red line) \cite{mpb-art}. At the bottom the real part
of two defect modes with $\epsilon_{def}=1$ and $\epsilon_{def}=30$
is shown.}
\label{fig:pointDefect} 
\end{figure}

In this section the quality of IM optimized Wannier functions is demonstrated
by using them as a basis set for photonic crystal defect structure
analysis. In figure \ref{fig:reprBS} the reconstructed tight-binding
band structure of a square lattice of dielectric rods in air (TM polarization)
is shown. The number of lattice sites taken into account in the nearest
neighbor approximations were increased successively. The deviation
of the reconstructed band structure from the original band structure
decreases by increasing the number of next neighbors taken into account.
By restricting to lattice sites separated by up to four lattice constants
the band structure is reproduced well except for small deviation at
higher bands and symmetry points. The slight mismatch for higher bands
is due to the fact that the higher band Wannier functions are not
as well localized as the lower ones.

\begin{figure}[t]
\centering \includegraphics[angle=-90, width=0.95\columnwidth]{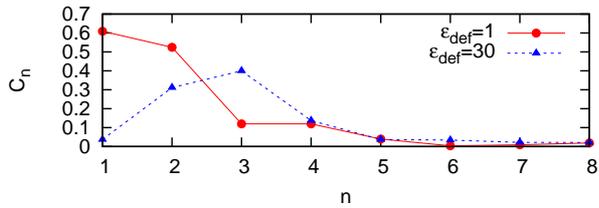} 

\caption[Band index contribution]{The band index contribution $C_{n}$ for the defect modes $\epsilon_{def}=1$
and $\epsilon_{def}=30$. Since the defect frequency lies inside the
band gap between the 1st and the 2nd band, only the lower band Wannier
functions contribute to the defect mode.}
\label{fig:bandContri} 
\end{figure}

We further calculated the modes and frequencies of a point defect
structures which consist of a single rod with deviating permittivity
$\epsilon_{def}$ at $\ve{R}_{def}$ in the Sq-D structure. The light
propagation inside the band gap between the first and the second band
is forbidden and the formation of a localized defect mode is possible.
We split up the total permittivity $\epsilon(\ve{r})=\epsilon_{p}(\ve{r})+\delta\epsilon(\ve{r}),$
into a periodic term $\epsilon_{p}(\ve{r})$ which corresponds to
the unperturbed system, and a defect term $\delta\epsilon(\ve{r})=(\epsilon_{def}-\epsilon_{p})\Theta(r-|\ve{R}_{def}-\ve{r}|)$.
An expansion of the $z$-component of the electric field in terms
of maximally localized Wannier functions leads to a generalized eigenvalue
problem with sparse matrices \cite{Busch2003}. The defect mode frequencies are obtained
from the solution of this sparse system (Fig. \ref{fig:pointDefect}).
The first eight Wannier functions have been taken into account. There
are monopole like defect modes for a defect rod permittivity of $\epsilon_{def}<12$
and doubly degenerated, dipole like modes for higher permittivities
in the defect rod $\epsilon_{def}>12$. The results are in complete
agreement with plane wave calculations \cite{mpb-art}. To analyze
contribution of the individual Wannier functions to the defect mode,
the band index contribution
\begin{equation}
C_{n}=\frac{1}{M}\sum_{\ve{R}}|\blfs{E}{n}{R}|^{2}
\end{equation}
where the normalization factor is given by $M=\sum_{n\ve{R}}|\blfs{E}{n}{R}|^{2}$
is shown in figure \ref{fig:bandContri} for the defect modes $\epsilon_{def}=1$
and $\epsilon_{def}=30$. One can see, that $C_{n}$ rapidly decrease
for higher band indices. Since the defect mode frequencies are in
between the 1st and the 2nd band, only the lower band Wannier functions
contribute to the defect modes. Therefore it is justified to cut of
the band index which reduces the numerical load.

\section{Conclusion\label{sec:Conclusion}}

The procedure to construct maximally localized Wannier functions by
Bloch phase optimization was analyzed for several two dimensional
photonic crystals for both fundamental polarizations, using two different 
locality measures. 
Although the stochastic, genetic algorithm is numerically too costly
for routine application, it has, as a global optimization method, 
provided us an important benchmark to judge under which conditions the 
faster and less memory intensive, but local conjugate gradient method 
finds the global optimum of a given locality measure. 
We found that the commonly used second moment locality measure has 
generically multiple extrema, which makes it difficult to construct 
maximally localized Wannier functions by local optimization techniques.
One may conjecture that this multiplicity results from the complex oscillatory
behavior of the (not yet optimized) Wannier functions at large distances
from the Wannier center, which makes the dominant contribution to the
second moment. This led us to propose a new locality measure which is 
controlled by the behavior close the Wannier center, the integrated 
modulus square measure. We showed numerically by comparison of 
conjugate gradient and genetic algorithm optimization that this measure 
does not feature multiple extrema and is, therefore, suitable for 
fast and efficient local optimization techniques, like the standard
conjugate gradient method. Because this result presumably originates 
from the local nature of the integrated modulus measure, it should hold 
generally, not only for two-dimensional systems, but also in three dimensions 
for photonic as well as electronic lattices.
 
We also presented and tested an analytical formula for the set of Bloch phases 
to be used as a starting point of the optimization process. 
This initial set of Bloch phases is suggested because, albeit it does 
not solve the optimization problem in general, it does generate 
maximally localized Wannier functions in special cases where the 
optimization problem can be solved analytically. 
We expect that these two main results may significantly increase the 
efficiency of the Wannier function approach for the description of 
defect structures in photonic lattices. 

This work was supported in part by the Deutsche
Forschungsgemeinschaft (FOR 557).

\end{document}